# Room-Temperature Spin-Valve Effect in Fe$_3$GaTe$_2$/MoS$_2$/Fe$_3$GaTe$_2$ 2D van der Waals Heterojunction Devices

*Wen Jin[1,2], Gaojie Zhang[1,2], Hao Wu[1,2,3,4], Li Yang[1,2], Wenfeng Zhang[1,2,3], Haixin Chang[1,2,3,4,\*]*

[1]Center for Joining and Electronic Packaging, State Key Laboratory of Material Processing and Die & Mold Technology, School of Materials Science and Engineering, Huazhong University of Science and Technology, Wuhan 430074, China.

[2]Shenzhen R&D Center of Huazhong University of Science and Technology (HUST), Shenzhen 518000, China.

[3]Institute for Quantum Science and Engineering, Huazhong University of Science and Technology, Wuhan 430074, China.

[4]Wuhan National High Magnetic Field Center, Huazhong University of Science and Technology, Wuhan 430074, China.

[\*]Corresponding author. E-mail: hxchang@hust.edu.cn

## Abstract

Spin-valve effect has been the focus of spintronics over the last decades due to its potential in many spintronic devices. Two-dimensional (2D) van der Waals (vdW) materials are highly expected to build the spin-valve heterojunction. However, the Curie temperatures (T$_C$) of the vdW ferromagnetic 2D crystals are mostly below room temperature (~30-220 K). It is very challenging to develop room temperature, ferromagnetic (FM) 2D crystals based spin-valve devices which are still not available to date. We report the first room temperature, FM 2D crystal based all-2D vdW Fe$_3$GaTe$_2$/MoS$_2$/Fe$_3$GaTe$_2$ spin valve devices. The Magnetoresistance (MR) of the all-devices is up to 15.89% at 2.3 K and 11.97% at 10 K, 4-30 times of MR from the spin valves of $Fe_3GeTe_2/MoS_2/Fe_3GeTe_2$ and conventional $NiFe/MoS_2/NiFe$. Typical spin valve effect shows strong dependence on $MoS_2$ spacer thickness in the vdW heterojunction. Importantly, the spin valve effect (0.31%) still robustly exists at 300 K with low working currents down to 10 nA (0.13 A/cm$^2$). The results provide a general vdW platform to room temperature, 2D FM crystals based 2D spin valve devices.

## 1. Introduction

Spin-valve effect has been the focus of spintronics over the last decades due to its application in magnetoresistance sensors[1-3], magnetic random-access memory[4,5] and read head for high density magnetic recording[6,7]. Typically, spin-valve devices are consisted of two ferromagnetic (FM) electrodes separated by a nonmagnetic spacer layer. Traditional ferromagnets such as Fe[8], NiFe[9], Co[10] or CoFeB[11] are extensively investigated as the FM electrodes of spin-valve devices. To obtain a well-defined interface regardless of lattice mismatch, two-dimensional (2D) van der Waals (vdW) materials are highly expected to build the spin-valve heterojunction due to their perfectly flat surface without dangling bonds[12].

Realizing room-temperature ferromagnetism in 2D vdW materials is essential for the practical application of 2D spin-valve devices in spintronics. However, theoretically, long-range ferromagnetism is hard to exist in 2D materials because of thermal fluctuations[13], while it can be stabilized by the excitation gap induced by magnetic anisotropy[14]. To date, progresses have been made in the research of intrinsic 2D ferromagnetic crystals, such as $CrI_3$[15], $Cr_2Ge_2Te_6$[14] and $Fe_3GeTe_2$[16]. However, the Curie temperatures ($T_C$) of these ferromagnetic 2D crystals are mostly below room temperature (~30-220 K), hindering the practical room temperature applications in next-generation spin-valve spintronic devices. Thus, it is very challenging to develop room temperature, FM 2D crystals based spin-valve devices which are still not available to date.

Very recently, the emergence of $Fe_3GaTe_2$ with above-room temperature Tc and robust large room-temperature perpendicular magnetic anisotropy (PMA)[17] makes the practical room-temperature application of 2D vdW FM crystals based spin-valve devices in spintronics possible. In this work, we report the first room temperature all-2D vdW $Fe_3GaTe_2$/$MoS_2$/$Fe_3GaTe_2$ spin valve devices with $Fe_3GaTe_2$ as the top and bottom FM electrodes and few layer $MoS_2$ as the spacer. Typical spin valve effect is observed in the heterojunction as the formation of Ohmic contacts at the $Fe_3GaTe_2$/$MoS_2$ interfaces according to linear current-voltage (I-V) curves. The MR of the all-2D vdW spin valve device is up to 15.89% at 2.3 K and still robustly exists (0.31%) at 300 K with low working currents down to 10 nA (0.13 A/cm$^2$) at both 2.3 K and 300 K.

## 2. Results and Discussion

The crystal structure of $Fe_3GaTe_2$ is schematically illustrated in Figure 1a, which is hexagonal that belongs to space group $P6_3/mmc$ ($a=b=3.9860$ Å, $c=16.2290$ Å, $α=β=90°$, $γ=120°$). In 2D $Fe_3GaTe_2$, $Fe_3Ga$ heterometallic slab is in the middle with Te atoms on the top and bottom sides, forming a typical sandwiched structure. The adjacent layers are connected by weak vdW force with an interlayer thickness of 0.78 nm. The magneto-transport properties are revealed by Anomalous Hall Effect (AHE) of a Hall device. As shown in Figure 1b, the thickness of the 2D $Fe_3GaTe_2$ crystal in Hall device is 16 nm. Figure 1e shows the temperature-dependent longitudinal resistance $R_{xx}$, as temperature decreases from 300 K to 2 K, and the $R_{xx}$ declines monotonously, implying the metallic characteristics of $Fe_3GaTe_2$. Meanwhile, the corresponding I-V curves at different temperature in the inset of Figure 1e is linear, which verifies the Ohmic contact between the electrodes and $Fe_3GaTe_2$. Figure 1c is the Anomalous Hall resistance ($R_{xy}$) of the device with an out-of-plane magnetic field at various temperatures. The large PMA is clearly indicated by the rectangular hysteresis loops and it shows the $T_C$ is up to 340 K, which are all similar to our previous report [17]. The coercivities ($H_C$) extracted from the AHE results (Figure 1d) are temperature-dependent, which decline as the temperature increases and that is a result of thermal fluctuation intensify. The above-room-temperature Tc, strong PMA and large $H_C$ verify the good quality of $Fe_3GaTe_2$ crystal, making it an ideal material for metallic ferromagnetic electrode of room-temperature spin valve.

The spin valve device is consisted of two FM electrodes ($Fe_3GaTe_2$) and a non-magnetic spacer layer (MoS$_2$), the diagram of the structure is shown in Figure 2a. MoS$_2$ is a semiconductor with tunable bandgap (1.8 eV in monolayer and 1.2 eV in bulk[18-20]), whereas its vertical conductivity is relatively poor due to the weak interlayer interaction[21] and thus making it appropriated for spacer layer in spin-valve devices. To avoid damage and contamination caused during the deposition of Cr/Au electrodes, we fabricated the metal electrodes first and the following procedure can be finished in the Ar-filled glove box. We applied four-terminal setup to measure the magnetoresistance (MR) so that the contact resistance of the device can be excluded. The magneto-transport measurements were carried out under a magnetic field with the direction perpendicular to the *ab* plane. To observe spin-valve effect, the switching fields of the two ferromagnetic electrodes should be different [22,23]. The switching fields are decided by the coercivity of Fe$_3$GaTe$_2$, which depends on the geometry and thickness of Fe$_3$GaTe$_2$ [24]. Thus, we selected Fe$_3$GaTe$_2$ with different geometry and thickness as the FM metal layer. The AFM image and the thickness profile of the device are shown in Figure S1 in Supporting Information, indicating the top and bottom Fe$_3$GaTe$_2$ 16.8 nm and 9.5 nm, respectively, and the middle few layer MoS$_2$ 4.5 nm. In addition, Raman spectra for the top and bottom Fe$_3$GaTe$_2$, MoS$_2$ and the heterojunction are shown in Figure S2 in Supporting Information, indicating the successful formation of the multilayer heterojunction. Moreover, the Raman signals for MoS$_2$ exhibit in-plane vibration mode $E^1_{2g}$ (~383 cm$^{-1}$) and out-of-plane $A_{1g}$ (~406 cm$^{-1}$), the difference between $E^1_{2g}$ and $A_{1g}$ is ~23 cm$^{-1}$, suggesting the multilayer of MoS$_2$[25,26], which is consistent with the AFM tests. Figure 2c is a typical MR curve with 1 μA bias current at 300 K. The perpendicular magnetic field was applied to sweep between -0.05 T and 0.05 T. When sweeping from -0.05 T to 0.05 T (blue square), the resistance increases abruptly at 50 Oe and sustains until B=200 Oe. As the applied magnetic field becomes larger, the resistance exhibited a sudden decrease. When sweeping backwards (red square), a similar sudden increase and a followed-by decrease appeared. The two distinct values of resistance are corresponding to the magnetization switching of the two $Fe_3GaTe_2$. To begin with, the magnetization direction of the two $Fe_3GaTe_2$ is parallel, corresponding to low-resistance. When reversing magnetization, the magnetization vector of the $Fe_3GaTe_2$ with small coercivity will turn over first, making the magnetization direction of the two $Fe_3GaTe_2$ become antiparallel. According to the optical image of the device shown in Figure 2b, the area of the heterojunction can be estimated as 7.75 μm$^2$, and thus the resistance-area product (RA) is 10.25 kΩ μm$^2$ at 300 K. The MR ratio can be obtained according to $MR = (R_{AP} - R_P)/R_P$, where $R_{AP}$ and $R_P$ refer to the antiparallel and parallel magnetic configurations of the two $Fe_3GaTe_2$, respectively. According to the equation, the MR ratio of $Fe_3GaTe_2$/$MoS_2$/$Fe_3GaTe_2$ heterojunction can be determined as 0.31% at 300 K (Figure 2c) and 15.89% at 2.3 K (see data and discussion below for Figure 4a), which is the first ferromagnetic 2D crystals based all 2D vdW room-temperature spin-valve device. The MR at 10 K (11.97%) is approximately 30 times and 16 times to the conventional spin valves of NiFe/$MoS_2$/NiFe (0.4%) and NiFe/Au/$MoS_2$/NiFe (0.73)[27] respectively, and 4 times to $Fe_3GeTe_2$/$MoS_2$/$Fe_3GeTe_2$ (3.1%)[28] spin valve in previous reports. Moreover, a reversed bias current was applied to the MR measurement (Figure 2d), the MR ratio shows little difference comparing that in Figure 2c, suggesting the interface in the fabricated spin valve is symmetric.

The electrical characteristic was investigated afterwards. Figure 3a depicts the temperature-dependent resistance, exhibiting an overall downward trend and indicating the metallic behavior. The vertical heterojunction resistance is dominated by $MoS_2$, which is a semiconductor while the junction shows a metallic behavior. This can be explained as a result of strong hybridization between the S atoms of $MoS_2$ and Fe atoms like that in $Fe_3GeTe_2$ heterostructures[29]. Note that there is an increase as the temperature below 50 K, which is due to the Kondo effect. The linear current-voltage (I-V) curves at various temperatures in **Figure 3b** suggest the Ohmic contact between $Fe_3GaTe_2$ and $MoS_2$. Besides, they also imply there is no tunnel barrier formed in the heterojunction and $MoS_2$ is acting as a conducting layer. **Figure 3c** shows the I-V curves tested in parallel ($I_P$) and antiparallel ($I_{AP}$) magnetization at 2.3 K, where the low- and high-resistance can be obtained, corresponding to the MR-B test.

To investigate the influence of the performing current on the performance of the spin valve, the MR measured at various current bias currents were carried out at 2.3 K and 300 K, respectively. As shown in Figure 4a and 4b, spin-valve effects are observed at bias currents ranging from 10 nA to 30 μA and maintain stable at low working currents down to 10 nA (0.13 $A/cm^2$) at both 2.3 K and 300 K. Note that the current intensity could have been lower but the test was limited by our physical property measurement system. This indicates that the spin valve can work as a low-power-consumption device at room-temperature, which has a great potential in room-temperature spintronics. Figure 4c and 4d are the current-dependent MR extracted from Figure 4a and 4b, respectively. At 2.3 K, when the bias current is smaller than 1 μA, the MR ratios can maintain at ~15.5%, suggesting that the spin valve can work with a wide range of bias currents. As the applied bias current increases exponentially, the MR almost decreases linearly, which might be corresponding to the higher energy electrons scattering at the interface of $Fe_3GeTe_2/MoS_2$[30]. At 300 K, the MR ratio shows little dependence on the bias current, we speculate that is because of the thermal fluctuation enhancement diminishes the electrons scattering[30]. To understand the influence of the $MoS_2$ thickness on the MR ratio of the spin valve, other devices with three different $MoS_2$ thicknesses were fabricated, as shown in Figure S4. Three representative devices were all tested at 2.3 K with 1 μA bias current and the MR shows a spacer layer thickness-dependent behavior. When the $MoS_2$ thickness is 8 nm, the MR is 2.4%. With the thickness of $MoS_2$ increase to 10 nm, the MR drops to 0.77% and finally down to 0.53% when the $MoS_2$ thickness is 17 nm.

The temperature dependence of the spin valve was further studied by measuring the MR ratios at various temperatures up to 310 K. The MR ratios are extracted from Figure 5a and shown in Figure 5b. The MR ratio monotonically decreases as the temperature increases and still retains clearly at room temperature, corresponding to the Tc of $Fe_3GeTe_2$. The temperature-dependence MR is closely related to the spin polarization, which can be described as $MR = \frac{P_1 P_2}{1 - P_1 P_2}$, according to Julliere model [31], where $P_1$ and $P_2$ refer to the spin polarization value of the top and bottom FM layer. In our spin valve device, the two FM layers are used by the same material, so we set $P_1 \approx P_2 = P$. Thus, the relationship between the calculated P and temperature is depicted in Figure 5c, the P value decreases from 27% to 4% as the temperature increases from 2.3 K to 300 K. The temperature dependence of P can be fitted by the Bloch's law[32] with the equation of $P(T) = P(0)(1 - \alpha T^{\frac{3}{2}})$, where P(0) refers to the spin polarization at 0 K and α is a materials dependent constant[33,34]. The α of our spin valve is $1.6 \times 10^{-4}$ K$^{-3/2}$, which is larger than that for Co ($1\text{-}6 \times 10^{-6}$ K$^{-3/2}$) and NiFe ($3\text{-}5 \times 10^{-5}$ K$^{-3/2}$)[35]. Generally, α is larger for surface compared to bulk due to the soften of surface exchange[36,37]. The larger α here maybe the result of interface scattering [38].

## 3. Conclusion

We report the first room temperature, ferromagnetic 2D crystals based all-2D vdW spin valve heterojunction devices. Typical spin valve effect shows strong spacer thickness dependence on $MoS_2$ in the vdW heterojunctions. The MR of the all-2D vdW spin valve device is up to 15.89% at 2.3 K and 11.97% at 10 K, 4-30 times of MR from the spin valves of $Fe_3GeTe_2/MoS_2/Fe_3GeTe_2$ and conventional $NiFe/MoS_2/NiFe$. Importantly, the spin valve effect still robustly exists at 300 K with low working currents down to 10 nA (0.13 A/cm$^2$). The results provide a general vdW platform to room temperature, 2D FM crystals based 2D spin valve devices.

**References**

[1] M. A. Khan, J. Sun, B. Li, A. Przybysz, J. Kosel, *Engineering Research Express* **2021**, 3, 022005.
[2] J.-G. Zhu, C. Park, *Mater. Today* **2006**, 9, 36.
[3] C. Ren, Q. Bayin, S. Feng, Y. Fu, X. Ma, J. Guo, *Biosens. Bioelectron.* **2020**, 165, 112340.
[4] B. Behin-Aein, D. Datta, S. Salahuddin, S. Datta, *Nat. Nanotechnol.* **2010**, 5, 266.
[5] B. Dieny, I. L. Prejbeanu, in *Introduction to Magnetic Random-Access Memory*, **2017**, p. 101.
[6] W.-H. Hsu, R. H. Victora, *Appl. Phys. Lett.* **2021**, 118.
[7] T. Nakatani, Z. Gao, K. Hono, *MRS Bull.* **2018**, 43, 106.
[8] K. Dolui, A. Narayan, I. Rungger, S. Sanvito, *Phys. Rev. B* **2014**, 90.
[9] M. Z. Iqbal, M. W. Iqbal, J. H. Lee, Y. S. Kim, S.-H. Chun, J. Eom, *Nano Res.* **2013**, 6, 373.
[10] J. Meng, J.-J. Chen, Y. Yan, D.-P. Yu, Z.-M. Liao, *Nanoscale* **2013**, 5, 8894.
[11] R. B. Morgunov, G. L. L'Vova, A. D. Talantsev, Y. Lu, X. Devaux, S. Migot, O. V. Koplak, O. S. Dmitriev, S. Mangin, *Thin Solid Films* **2017**, 640, 8.
[12] M. D. Siao, W. C. Shen, R. S. Chen, Z. W. Chang, M. C. Shih, Y. P. Chiu, C. M. Cheng, *Nat. Commun.* **2018**, 9, 1442.
[13] N. D. Mermin, H. Wagner, *Phys. Rev. Lett.* **1966**, 17, 1133.
[14] C. Gong, L. Li, Z. Li, H. Ji, A. Stern, Y. Xia, T. Cao, W. Bao, C. Wang, Y. Wang, Z. Q. Qiu, R. J. Cava, S. G. Louie, J. Xia, X. Zhang, *Nature* **2017**, 546, 265.
[15] B. Huang, G. Clark, E. Navarro-Moratalla, D. R. Klein, R. Cheng, K. L. Seyler, D. Zhong, E. Schmidgall, M. A. McGuire, D. H. Cobden, W. Yao, D. Xiao, P. Jarillo-Herrero, X. Xu, *Nature* **2017**, 546, 270.
[16] Y. Deng, Y. Yu, Y. Song, J. Zhang, N. Z. Wang, Z. Sun, Y. Yi, Y. Z. Wu, S. Wu, J. Zhu, J. Wang, X. H. Chen, Y. Zhang, *Nature* **2018**, 563, 94.[17] G. Zhang, F. Guo, H. Wu, X. Wen, L. Yang, W. Jin, W. Zhang, H. Chang, *Nat. Commun.* **2022**, 13, 5067.
[18] B. Radisavljevic, A. Kis, *Nat. Mater.* **2013**, 12, 815.
[19] S. Chuang, C. Battaglia, A. Azcatl, S. McDonnell, J. S. Kang, X. Yin, M. Tosun, R. Kapadia, H. Fang, R. M. Wallace, A. Javey, *Nano Lett*. **2014**, 14, 1337.
[20] K. F. Mak, C. Lee, J. Hone, J. Shan, T. F. Heinz, *Phys. Rev. Lett.* **2010**, 105, 136805.
[21] M. Xu, T. Liang, M. Shi, H. Chen, *Chem. Rev.* **2013**, 113, 3766.[22] I. Žutić, J. Fabian, S. Das Sarma, *Rev. Mod. Phys.* **2004**, 76, 323.
[23] J. M. D. Coey, *Magnetism and Magnetic Materials*, Cambridge University Press, Cambridge **2010**.
[24] C. Tan, J. Lee, S.-G. Jung, T. Park, S. Albarakati, J. Partridge, M. R. Field, D. G. McCulloch, L. Wang, C. Lee, *Nat. Commun.* **2018**, 9, 1554
[25] H. Li, Q. Zhang, C. C. R. Yap, B. K. Tay, T. H. T. Edwin, A. Olivier, D. Baillargeat, *Adv. Funct. Mater.* **2012**, 22, 1385.
[26] C. Lee, H. Yan, L. E. Brus, T. F. Heinz, J. Hone, S. Ryu, *ACS Nano* **2010**, 4, 2695.[27] W. Wang, A. Narayan, L. Tang, K. Dolui, Y. Liu, X. Yuan, Y. Jin, Y. Wu, I.

Rungger, S. Sanvito, F. Xiu, *Nano Lett.* **2015**, 15, 5261.

[28] H. Lin, F. Yan, C. Hu, Q. Lv, W. Zhu, Z. Wang, Z. Wei, K. Chang, K. Wang, *ACS Appl. Mater. Interfaces* **2020**, 12, 43921.

[29] K. Dolui, A. Narayan, I. Rungger, S. Sanvito, *Phys. Rev. B* **2014**, 90, 041401.[30] J. Zhang, R. M. White, *J. Appl. Phys.* **1998**, 83, 6512.[31] M. Julliere, *Phys. Lett. A* **1975**, 54, 225.

[32] R. Sharif, S. Shamaila, F. Shaheen, J. Y. Chen, M. Khaleeq-ur-Rahman, K. Hussain, *Appl. Phys. Lett.* **2013**, 102.[33] C. H. Shang, J. Nowak, R. Jansen, J. S. Moodera, *Phys. Rev. B* **1998**, 58, R2917.

[34] J. Mathon, S. B. Ahmad, *Phys. Rev. B* **1988**, 37, 660.

[35] C. H. Shang, J. Nowak, R. Jansen, J. S. Moodera, *Phys. Rev. B* **1998**, 58, R2917.

[36] D. L. Mills, A. A. Maradudin, *J. Phys. Chem. Solids* **1967**, 28, 1855.

[37] J. Mathon, S. B. Ahmad, *Phys. Rev. B* **1988**, 37, 660.

[38] J. Unguris, D. T. Pierce, R. J. Celotta, *Phys. Rev. B* **1984**, 29, 1381.

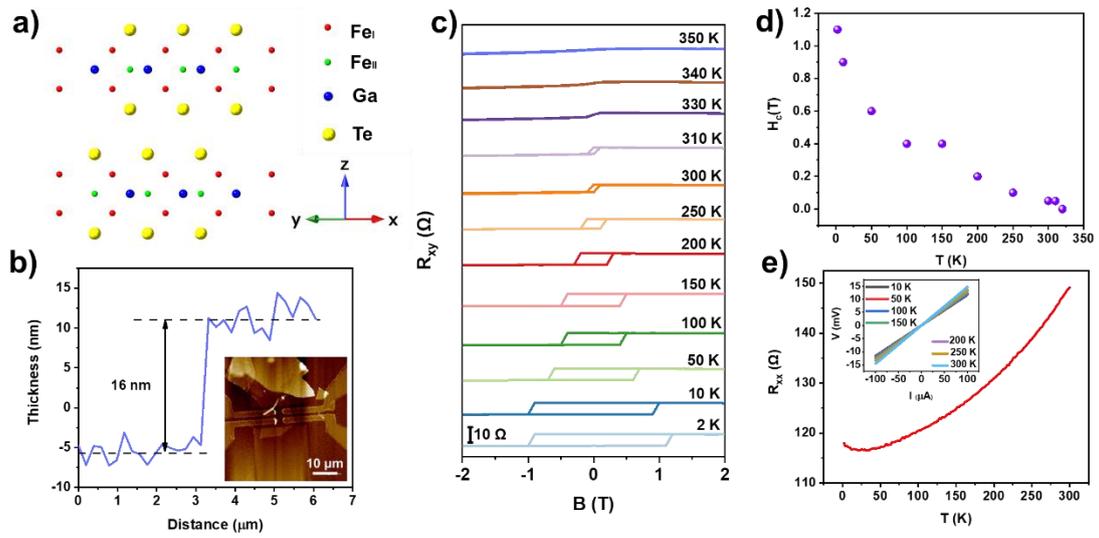

**Figure 1.** Crystal structure and magneto-transport characterization of the vdW layered Fe$_3$GaTe$_2$ single crystals. a) Front view of the crystal structure of Fe$_3$GaTe$_2$. b) AFM height profile and topography (inset) of the Fe$_3$GaTe$_2$ Hall device. c) Hall resistance ($R_{xy}$) at different temperatures from 2 K to 350 K. d) Extracted coercivities as a function of temperature. e) Temperature dependence of the longitudinal resistance ($R_{xx}$). Inset: I-V curves at different temperatures.

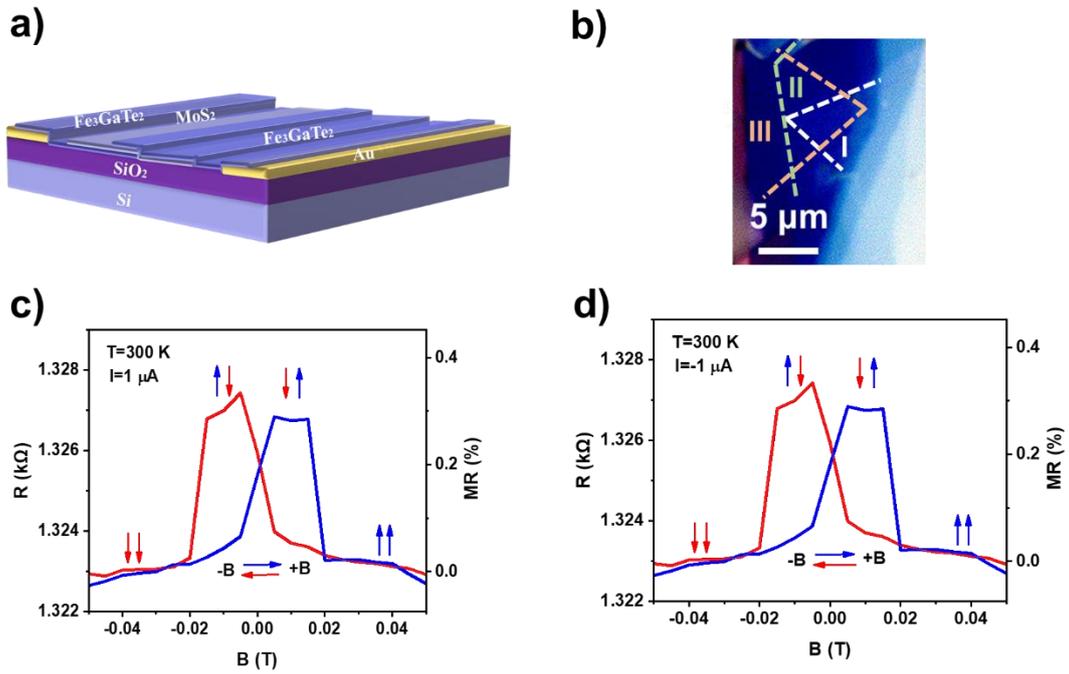

**Figure 2.** Spin valve device characterization and magneto-transport properties. a) The schematic diagram of the $Fe_3GaTe_2$/$MoS_2$/$Fe_3GaTe_2$ heterojunction spin valve device. b) Optical image of the device, regions I, II and III represent bottom $Fe_3GaTe_2$, MoS2, and top $Fe_3GaTe_2$, respectively. c,d) Resistance and MR vs perpendicular magnetic field at 300 K with a bias current of 1 µA and -1 µA. The arrows represent the magnetizations alignment directions of $Fe_3GaTe_2$.

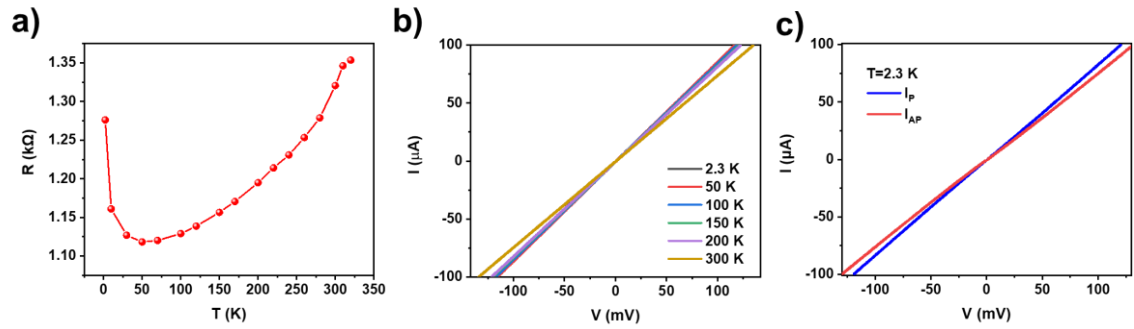

**Figure 3.** Electrical and magneto-transport properties of the $Fe_3GaTe_2/MoS_2/Fe_3GaTe_2$ heterojunction spin valve device. a) Resistance vs temperature of the spin valve device. b) I-V curves at different temperatures. c) I-V curves measured at parallel ($I_P$) and antiparallel ($I_{AP}$) magnetization at 2.3 K.

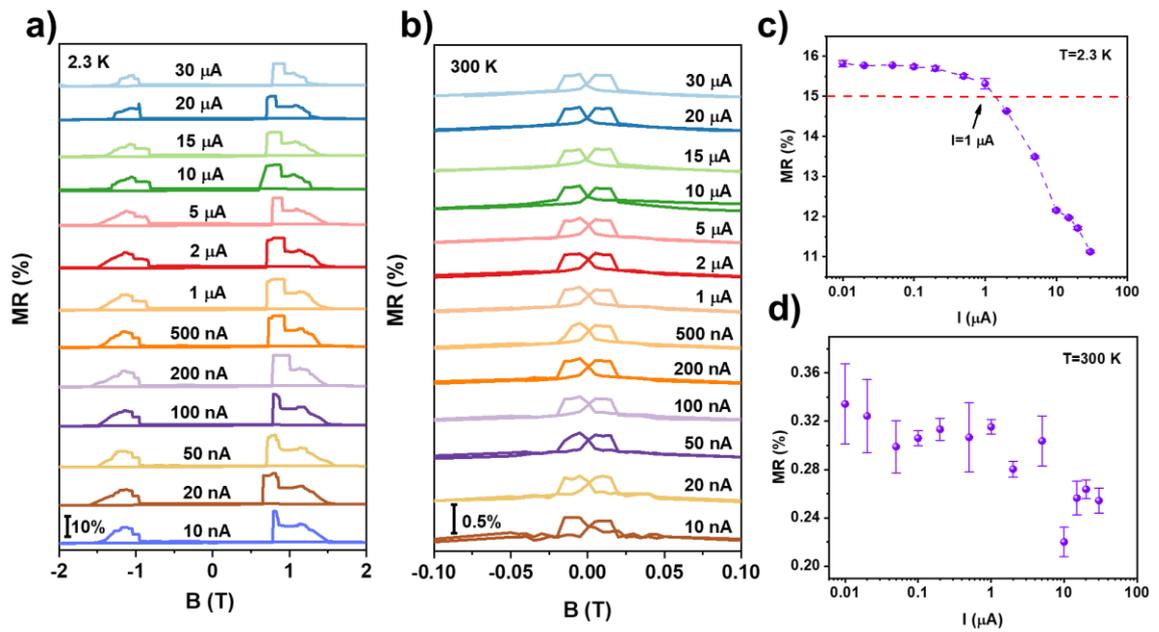

**Figure 4.** Bias current dependence of spin valve effect for $Fe_3GaTe_2$/$MoS_2$/$Fe_3GaTe_2$ device. a,b) MR curves measured at different bias current from 10 nA to 30 μA at (a) 2.3 K and (b) 300 K, respectively. c,d) Extracted MR ratios of as a function of bias current at (c) 2.3 K and (d) 300 K, respectively.

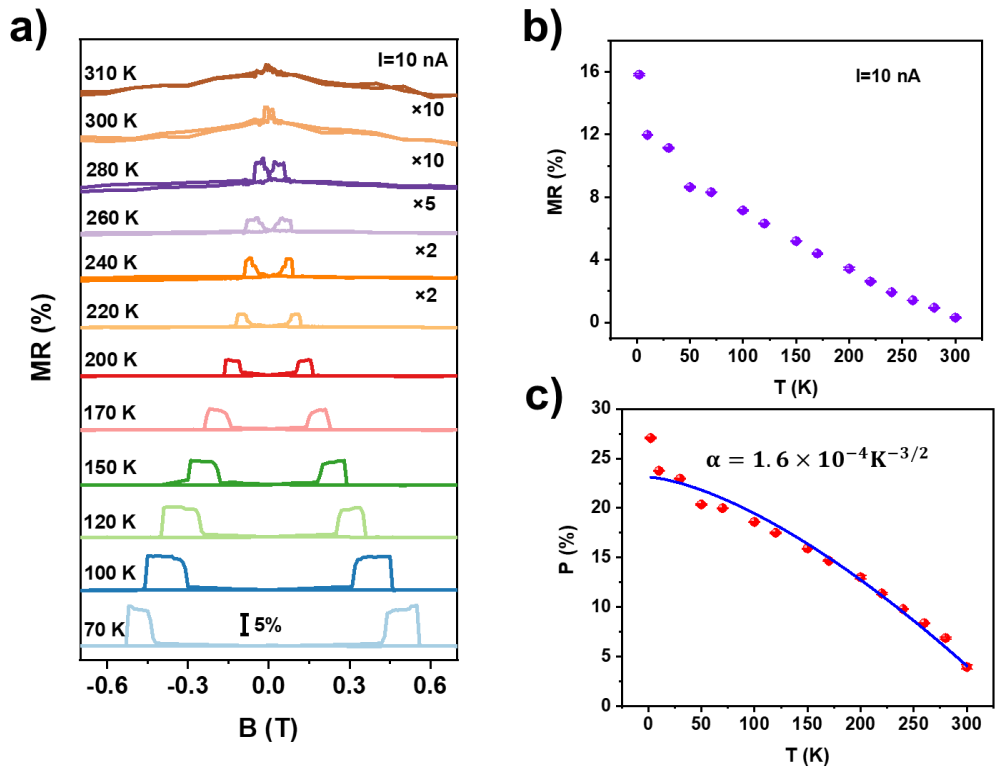

**Figure 5.** Temperature dependence of spin valve effect for $Fe_3GaTe_2/MoS_2/Fe_3GaTe_2$ device. a) MR curves measured at different temperatures from 70 K to 310 K with a fixed bias current of 10 nA. b) Extracted MR ratios of as a function of temperature. c) Spin polarization as a function of temperature. The blue line is the fitting curve according to Bloch's law.